\documentclass[twocolumn,amsmath,amssymb]{revtex4-1}

\usepackage{graphicx}
\usepackage{bm}

\begin{document}

\title{Long-range interactions between polar alkali-metal diatoms in external electric fields}

\author{M. Lepers$^{1}$, R. Vexiau$^{1}$, M. Aymar$^{1}$, N. Bouloufa-Maafa$^{1,2}$ and O. Dulieu$^{1}$}
\affiliation{${}^{1}$Laboratoire Aim\'e Cotton, CNRS/Univ.~Paris-Sud/ENS-Cachan, B\^at.~505, Campus d'Orsay, 91405 Orsay, France}
\email{maxence.lepers@u-psud.fr}
\affiliation{${}^{2}$Universit\'e Cergy-Pontoise, 95000 Cergy-Pontoise, France}

\date{\today}

\begin{abstract}
We computed the long-range interactions between two identical polar bialkali molecules in their rovibronic ground level, for all ten species involving Li, Na, K, Rb and Cs, using accurate quantum chemistry  results combined with available spectroscopic data. Huge van der Waals interaction is found for eight species in free space. The competition of the van der Waals interaction with the dipole-dipole interaction induced by an external electric field parallel or perpendicular to the intermolecular axis is investigated by varying the electric field magnitude and the intermolecular distance. Our calculations predict a regime with the mutual orientation of the two molecules but with no preferential direction in the laboratory frame. A mechanism for the stimulated one-photon radiative association of a pair of ultracold polar molecules into ultracold tetramers is proposed, which would open the way towards the optical manipulation of ultracold polyatomic molecules.
\end{abstract}

\maketitle

\section{\label{intro} Introduction}

The dynamics of ultracold quantum gases composed of atoms or molecules with extremely low translational energy $E_{t}/k_B \ll$~1~millikelvin is dominated by the long-range mutual interactions between particles. Such gases are nowadays routinely produced in various laboratories worldwide and many applications are foreseen \cite{carr2009,dulieu2009}. When they are trapped in external potentials created by electromagnetic fields, they offer unique opportunities to study fundamental few-body dynamics in atomic and molecular physics \cite{blume2012}. The unprecedented capability to simultaneously control the internal and external degrees of freedom of the particles also opens the way to the quantum simulation of Hamiltonians describing many-body physical phenomena like low-temperature Fermi fluids or artificial gauge fields \cite{bloch2012, galitski2013}. When the particles possess an intrinsic magnetic or electric dipole moment they interact through strong long-range anisotropic forces, \textit{i.e.}~depending on their mutual orientation, which strongly modifies the dynamics of the quantum gas \cite{lahaye2009,pasquiou2012} and enhances stereochemical properties of ultracold bimolecular reactions \cite{demiranda2011}.

The recent production of ultracold heteronuclear alkali-metal dimers in their lowest rovibronic \cite{ni2008, deiglmayr2008} and hyperfine level \cite{ospelkaus2010b} stimulates many studies in this perspective. The permanent electric dipole moment (PEDM) $d_0$ of such (polar) molecules in their own frame allows for manipulating them with static electric fields \cite{avdeenkov2002, avdeenkov2003} and electromagnetic fields \cite{micheli2007, holmegaard2009, demiranda2011, nielsen2012}. Such studies require a detailed modeling of the molecule-molecule long-range interactions inside the quantum gas with or without the presence of external fields \cite{byrd2012a, byrd2012b}. The most spectacular experimental achievements on ultracold dipolar molecular gases have been performed on KRb molecules \cite{ni2008, ospelkaus2010b, ospelkaus2010a, demiranda2011} which motivated a wealth of theoretical investigations on this species \cite{julienne2009, idziaszek2010a, quemener2010a, quemener2010b, byrd2012a}. An accurate description of long-range interactions involving the other heteronuclear alkali diatoms is strongly needed since they draw rising attention \cite{stwalley2010, zabawa2010, debatin2011, quemener2011a, julienne2011, takekoshi2012, wu2012, repp2013,tung2013}. Considering two identical polar molecules at large distances $R$ (in atomic units 1 $\mathrm{a.u.}\equiv a_0=0.0529133$ nm) between their individual center-of-mass (c-o-m) connected by the $z$ axis and with polar angles $(\theta_{i},\phi_{i})$, $i=1,2$ with respect to the $z$ axis, their mutual long-range dipole-dipole interaction is conveniently written in the coordinate system associated with the tetramer (T-CS) based on the $z$ axis
\begin{equation}
\mathrm{V}_\mathrm{dd}(R) = -\frac{d_0^{2}}{R^{3}} \left( 2\cos\theta_1 \cos\theta_2 - \sin\theta_1 \sin\theta_2 \cos(\phi_2-\phi_1) \right).
\label{eq:Vdd}
\end{equation}
(The atomic units of energy, where 1~$\mathrm{a.u.}\equiv 2\times\mathrm{Ryd}=219475$~cm$^{-1}$, will be used throughout the paper except otherwise stated). In the range of distances investigated here ($R\gtrsim 30$~a.u.), it is easy to check that the energy of the mechanical rotation between the two molecules, and thus the corresponding rotational couplings, are small compared to the van der Waals (vdW) interaction and to the rotational energy of the individual molecules, so that the main features of the system can be captured in the T-CS.

In this paper, we compute the long-range interactions between two identical bosonic polar bi-alkali ground-state molecules both in free space (Secs.~\ref{sec:C6} and \ref{sec:multi-chan}) and as a function of the magnitude of an external electric field parallel or perpendicular to the $z$ axis (Sec.~\ref{sec:Efield}). We use the stationary perturbation theory as in our previous investigations on atom-molecule systems \cite{lepers2010, lepers2011a, lepers2011b, lepers2011c, lepers2012}. In free space, we found that the vdW interaction varying as $-C_6/R^6$ is characterized by $C_6$ coefficients that are three orders of magnitude larger than those for alkali atoms, in agreement with the results of Ref.~\cite{zuchowski2013} which uses a different method. The isotropic vdW interaction competes with the expected anisotropic dipole-dipole interaction induced by one molecule on the other, or by an external electric field $\mathcal{E}$ \cite{micheli2007, julienne2011}. Among the ten species built from Li, Na, K, Rb, and Cs atoms, we show that the interactions for KRb and LiNa -- which posses the smallest PEDMs of the series, \textit{i.e.}~0.56 and 0.61 Debye respectively \cite{aymar2005} -- behave differently from species with a larger PEDM, ranging from 1.2 Debye for RbCs to 5.5 Debye for LiCs \cite{aymar2005}. For the latter molecules, our calculations predict for critical values of $R$ and $\mathcal{E}$ the mutual orientation of the two molecules but with no preferential direction in the lab frame, \textit{i.e.}~no anisotropy of their interaction. Our work complements the recent one by Byrd \textit{et al.} \cite{byrd2012b} where the authors investigated the long-range interaction between two bi-alkali ground-state polar molecules aligned by a strong external electric field in the head-to-tail or side-by-side configurations in the lab frame. Section \ref{sec:align} presents a prospective discussion about the possibility to create ultracold ground-state polar tetramers by stimulated one-photon radiative association of a pair of ultracold polar bialkali molecules. Finally Section \ref{sec:Conclu} contains more general conclusions and prospects.

\section{Calculation of $C_6$ coefficients in free space \label{sec:C6}}

In the lowest rovibrational level ($v=0, j=0$ for each molecule, hereafter referred to as the $|0\rangle$ level) of their ground electronic state $X^{1}\Sigma_{g}^{+}$, the molecules have no PEDM in the T-CS. Their interaction energy $V_{j_1j_2}(R)$ is determined by the operator $\hat{\mathrm{V}}_\mathrm{dd}$ taken at the second order of the perturbation expansion, \textit{i.e.} $-C_{6}/R^{6}$. The $C_6$ coefficients are expressed in terms of dynamic dipole polarizabilities at imaginary frequencies \cite{derevianko2010}
\begin{equation}
C_6 = \frac{3}{\pi} \int_0^{+\infty} d\omega \left[\alpha(i\omega)\right]^2 ,
\label{eq:C6-pola}
\end{equation}
where $\alpha(i\omega)$ is the isotropic polarizability of the rovibronic ground level $|0\rangle$, expressed as a sum over all excites levels $|n\rangle$ accessible by dipolar transition
\begin{equation}
\alpha(i\omega) = \frac{2}{3}\sum_{n\neq0} \frac{\omega_{n0}d_{n0}^{2}}{\omega_{n0}^{2}+\omega^{2}}.
\label{eq:pola-im}
\end{equation}
In Eq.~(\ref{eq:pola-im}), $\omega_{n0}$ and $d_{n0}$ are respectively the transition energies and transition dipole moments  between $|n\rangle$ and $|0\rangle$, which were extracted from a combination of accurate semi-empirical potential energy curves (PECs) and electronic transition dipole moments (TDMs) computed in our group \cite{aymar2005, aymar2007}, and available spectroscopic PECs (Table \ref{tab:MolData}). The wave functions for the rovibrational levels and for the dissociation continua were computed using our code based on Mapped Fourier Grid Hamiltonian method \cite{kokoouline1999}. The details of our calculations are given in Ref.~\cite{vexiau_phd_2012} and in a forthcoming paper.

\begin{table}
\begin{ruledtabular}
\begin{tabular}{rrrr}
            Molecule &                Ref. &  $d_0$ &       $B_0$ \\
                     &                     & (a.u.) & (cm$^{-1}$) \\
\hline
 $^{23}$Na$^{133}$Cs & \cite{docenko2004a} &  1.845 &       0.058 \\
    $^7$Li$^{133}$Cs & \cite{staanum2007}  &  2.201 &       0.187 \\
  $^{23}$Na$^{87}$Rb & \cite{docenko2004b} &  1.304 &       0.070 \\
     $^7$Li$^{87}$Rb & \cite{ivanova2011}  &  1.645 &       0.215 \\
      $^7$Li$^{39}$K & \cite{tiemann2009}  &  1.410 &       0.256 \\
   $^{23}$Na$^{39}$K & \cite{gerdes2008}   &  1.095 &       0.095 \\
  $^{39}$K$^{133}$Cs & \cite{ferber2009}   &  0.724 &       0.030 \\
 $^{87}$Rb$^{133}$Cs & \cite{docenko2011}  &  0.490 &       0.016 \\
\hline
   $^{39}$K$^{87}$Rb & \cite{pashov2007}   &  0.242 &       0.038 \\
     $^7$Li$^{23}$Na & \cite{fellows1989}  &  0.223 &       0.374 \\
\end{tabular}
\end{ruledtabular}
\caption{Permanent dipole moment $d_0$ (1~a.u.= 2.541 580 59 Debye) and rotational constant $B_0$  for each ground-state molecule in their $v=0$ level, obtained after averaging on the $v=0$ radial wave function related to the experimental potential energy curve given in the reference of the second column.}
\label{tab:MolData}
\end{table}

\begin{figure}
\includegraphics[width=8cm]{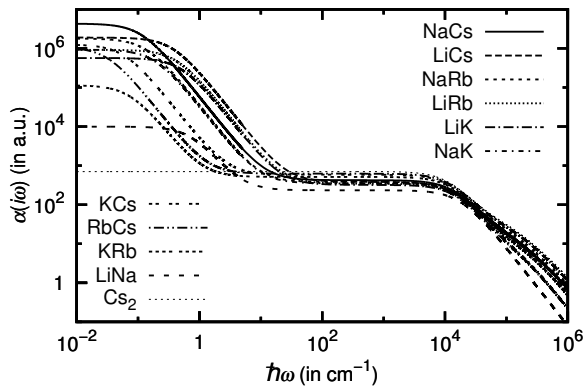}
\caption{Dynamic isotropic dipole polarizabilities as functions of the imaginary frequency for the ten heteronuclear bialkali molecules. We also plotted the polarizability of the homonuclear molecule Cs$_2$ as a reference \cite{lepers2011a}.}
\label{fig:Pola-Im}
\end{figure}

For the ten bialkali heteronuclear species, the polarizability $\alpha(i\omega)$, plotted on Fig.~\ref{fig:Pola-Im} in log-log scale, shows two plateaus. The first one, located in the low-frequency region ($\hbar \omega \approx 0.01-1$~cm$^{-1}$), is due to the predominance of the purely rotational transition $(v=0,j=0)\to (v=0,j=1)$ within the lowest electronic state $X$ in Eq.~(\ref{eq:pola-im}), over all other transitions toward ground state excited vibrational levels. The height of this plateau is strikingly different from one molecule to another, as it is shown by its approximate expression considering only the main transition $\alpha(0)=d_0^2/3B_0$ where the rotational constant $B_0$ of the $X^{1}\Sigma_{g}^{+}$ $v=0$ level and $d_0$ are expressed in atomic units. Once $\hbar \omega$ exceeds this transition energy, $\alpha(i\omega)$ varies as $\omega ^{-2}$ [Eq.~(\ref{eq:pola-im})] to reach the next plateau corresponding to the contribution of the transitions toward levels of the excited electronic states. This plateau extends up to $\hbar \omega \approx 10^4$~cm$^{-1}$ for all species which reflects (i) that the main contribution to the polarizability comes from the lowest excited electronic states, and (ii) that all species have similar electronic excitation energies. 
Then the polarizability varies as $\omega^{-2}$ up to $\hbar \omega \approx 10^5$~cm$^{-1}$, where the influence of the core electrons to the polarizability produces different slopes for $\alpha$.
Beyond this range, the polarizability decreases again as $\omega ^{-2}$. 

The existence of those two distinct plateaus allows us to express the $C_6$ coefficient as the sum of three terms \cite{quemener2011a}
\begin{equation}
C_6=C_{6}^{\mathrm{g}} + C_{6}^{\mathrm{e}} + C_{6}^{\mathrm{g-e}},
\label{eq:C6}
\end{equation}
where $C_6^g$ denotes the contribution of the purely rotational transition inside the ground electronic state (if we neglect transitions to higher vibrational levels), $C_6^e$ denotes the contribution of transitions to electronically-excited states,  and where $C_{6}^{\mathrm{g-e}}$ is a crossed contribution. Our results for all those contributions and for the ten heteronuclear molecules, are given in Table \ref{tab:C6}, and they are compared to available literature values. Just like the dynamic polarizabilities the $C_6$ coefficients vary dramatically from one molecule to another, ranging from a few thousands atomic units for LiNa and KRb, to a few millions for NaCs. Such differences ensue from the ground-state contribution, which can be expressed to a very good approximation as $C_{6}^{\mathrm{g}}\approx {d_0^{4}}/{6B_0}$. Eight molecules out of ten, which are characterized by a strong PEDM, interact through a huge van der Waals coefficient $C_6$, \textit{i.e.} larger than $10^5$ a.u.. The $C_{6}^{\mathrm{e}}$ values are comparable to those for homonuclear molecules \cite{byrd2011}, while the crossed terms $C_{6}^{\mathrm{g-e}}$ are always very weak.

\begin{table}
\begin{ruledtabular}
\begin{tabular}{rrrrrrrrrr}

           Molecule &                    Source &   $C_6$ & $C_6^g$ & $C_6^e$ & $C_6^{g-e}$ & $C_6^g/C_6$ \\
                    &                           &  (a.u.) &  (a.u.) &  (a.u.) &      (a.u.) &        (\%) \\
\hline
$^{23}$Na$^{133}$Cs &                 This work & 7323100 & 7311100 &    9198 &        2800 &        99.8 \\
                    & Ref.~\cite{zuchowski2013} & 6946696 & 6932958 &   10822 &        2916 &        99.8 \\
                    &     Ref.~\cite{byrd2012b} &         &         &    9453 &        2877 &
\vspace{6pt} \\
   $^7$Li$^{133}$Cs &                 This work & 4585400 & 4574400 &    7407 &        3600 &        99.8 \\
                    & Ref.~\cite{zuchowski2013} & 3409406 & 3397216 &    8670 &        3520 &        99.6 \\
                    &     Ref.~\cite{byrd2012b} &         &         &    7700 &        2920 & \\
                    & Ref.~\cite{quemener2011a} & 3840000 & 3830000 &    7712 &        3460 &        99.7
\vspace{6pt} \\
 $^{23}$Na$^{87}$Rb &                 This work & 1524900 & 1515800 &    7846 &        1200 &        99.4 \\
                    & Ref.~\cite{zuchowski2013} & 1507089 & 1497080 &    8696 &        1313 &        99.3 \\
                    &     Ref.~\cite{byrd2012b} &         &         &    7688 &         992 &
\vspace{6pt} \\
    $^7$Li$^{87}$Rb &                 This work & 1252300 & 1244205 &    6314 &        1800 &        99.4 \\
                    & Ref.~\cite{zuchowski2013} &  884705 &  876031 &    6963 &        1711 &        99.0 \\
                    &     Ref.~\cite{byrd2012b} &         &         &    6193 &        1061 & \\
                    & Ref.~\cite{quemener2011a} & 1070000 & 1070000 &    6323 &        1754 &     $>99.0$\vspace{6pt} \\
     $^7$Li$^{39}$K &                 This work &  570190 &  563500 &    5489 &        1200 &        98.8 \\
                    & Ref.~\cite{zuchowski2013} &  411682 &  404491 &    6024 &        1167 &        99.3 \\
                    &     Ref.~\cite{byrd2012b} &         &         &    5982 &        1261 & \\
                    & Ref.~\cite{quemener2011a} &  524000 &  517000 &    6269 &        1241 &        98.7
\vspace{6pt} \\
  $^{23}$Na$^{39}$K &                 This work &  561070 &  553520 &    6732 &         800 &        98.7 \\
                    & Ref.~\cite{zuchowski2013} &  516606 &  508325 &    7461 &         820 &        98.4 \\
                    &     Ref.~\cite{byrd2012b} &         &         &    6818 &         959 &
\vspace{6pt} \\
 $^{39}$K$^{133}$Cs &                 This work &  345740 &  329510 &   15619 &         611 &        95.3 \\
                    & Ref.~\cite{zuchowski2013} &  469120 &  450681 &   17716 &         723 &        96.1 \\
                    &     Ref.~\cite{byrd2012b} &         &         &   16570 &         690 &
\vspace{6pt} \\
$^{87}$Rb$^{133}$Cs &                 This work &  147260 &  129250 &   17707 &          53 &        87.8 \\
                    & Ref.~\cite{zuchowski2013} &  180982 &  160336 &   20301 &         345 &        88.6 \\
                    &     Ref.~\cite{byrd2012b} &         &         &   18840 &         370 &
\\
\hline
  $^{39}$K$^{87}$Rb &                 This work &   15972 &    3336 &   12576 &          60 &        20.9 \\
                    & Ref.~\cite{zuchowski2013} &   17720 &    3456 &   14202 &          62 &        19.5 \\
                    &     Ref.~\cite{byrd2012b} &         &         &   13490 &          50 &
\vspace{6pt} \\
    $^7$Li$^{23}$Na &                 This work &    3583 &     241 &    3321 &          21 &         6.7 \\
                    & Ref.~\cite{zuchowski2013} &    3709 &     110 &    3582 &          17 &         3.0 \\
                    &     Ref.~\cite{byrd2012b} &         &         &    3279 &          10 & \\
                    & Ref.~\cite{quemener2011a} &    3880 &     186 &    3673 &          21 &         4.8 \\
%
\end{tabular}
\end{ruledtabular}
\caption{The different contributions to the $C_6$ coefficient (see Eq.~(\ref{eq:C6})) between two identical ground-state bialkali heteronuclear molecules in their lowest rovibrational level. In Ref.~\cite{zuchowski2013}, $C_6^g$, $C_6^e$ and $C_6^{g-e}$ are denoted $C_6^\mathrm{rot}$, $C_6^\mathrm{disp}$ and $C_6^\mathrm{ind}$ respectively; in \cite{quemener2011a}, our $C_6^{g-e}$ is called $C_6^{inf}$; and in \cite{byrd2012b}, our $C_6^e$ and $C_6^{g-e}$ are $W_{6000}^{(2,\mathrm{DISP})}$ and $W_{6000}^{(2)}-W_{6000}^{(2,\mathrm{DIS})}$ respectively. In addition Ref.~\cite{kotochigova2010a} gives $C_6=16133$ a.u. and 142129 a.u.~for KRb and RbCs respectively; in \cite{buchachenko2012}, $C_{6,00}=13706$ a.u.~for KRb is similar to our $C_6^e+C_6^{g-e}$. The last column illustrate the existence of two classes of molecules as commented in the text.}
\label{tab:C6}
\end{table}

%
%

Table \ref{tab:C6} shows two trends: our results for $C_6^g$ are systematlcally larger than the other available values (except for KRb, KCs and RbCs from \cite{zuchowski2013}), while the contrary is visible for $C_6^e$. Such discrepancies probably come the sensitivity of the $C_6$ coefficients which scale as the fourth power of the permanent and transition dipole moments. In our calculations we use transition energies and dipole moments, averaged over rovibrational wave functions, while the other studies are done at the equilibrium distances. Those rovibrational wave functions are calculated from experimental PECs when possible. Our underestimation of the $C_6^e$ coefficients may be due to the slow convergence of the polarizability (Eq.~(\ref{eq:pola-im})) with respect to the electronically-excited states (and also ionization continuum), which are not all included in our calculations. The discrepancy of the $C_6^e$ may also be related to the overestimation of the static polarizability, in particular in Ref.~\cite{zuchowski2013} (see Tab.~1 and the corresponding discussion), which is inherent to the method used by the authors. Note that the static polarizabilities obtained with the present method differ by less than 1 \% from the values of Ref.~\cite{deiglmayr2008} obtained with a finite-field method. One possible way to discriminate the different theoretical calculations could be to measure collision rates, which scale for instance as $C_6^{3/4}$ for identical fermionic molecules \cite{quemener2011a}.

\section{Multi-channel calculation in free space
\label{sec:multi-chan}}

We consider in the following the eight molecules for which $C_{6}^{\mathrm{g}}$ is dominant (all but KRb and LiNa) as dipolar rotators characterized by $d_0$ and $B_0$, ignoring the influence of the electronically-excited states in their interaction.

The above single-channel description is valid down to distances $R^* = \left({d_0^{2}}/{B_0}\right)^{1/3}$ such that ${C_{6}}/{R^{*6}} \approx 2B_0$ where the dipole-dipole interaction couples the $(j_i=0)$ and $(j_i=1)$ levels \cite{micheli2007}. This is easily shown using an analytical model including the two channels $(j_1,j_2)=(0,0),(1,1)$, namely the basis states $|j_{1}m_{1};j_{2}m_{2}\rangle=|00;00\rangle$, $|10;10\rangle$ and $|1\pm 1;1\mp 1\rangle$, with $m_i$ the projection of $\vec{j}_i$ on $z$. Defining dimensionless energies $\bar{V}=V/B_{0}$ and distances $\bar{R}=R/R^*$, the lowest potential energy curve $\bar{V}_{0,0}(\bar{R})$ reads
\begin{equation}
\bar{V}_{0,0}(\bar{R}) \approx 2-2\sqrt{1+\frac{1}{6\bar{R}^{6}}}\,.
\label{eq:V-2chan}
\end{equation}
Thus $R^{*}$ is the distance where the variation of $\bar{V}_{0,0}$ suddenly changes from a variation in $\bar{R}^{-6}$ to $\bar{R}^{-3}$ due to the coupling with higher channels, inducing the mutual alignment of the molecules. As previously found \cite{julienne2011}, the values of $R^{*}$ are about two times smaller than the vdW length $R_{vdW}$ (Table \ref{tab:GrandCarac}) where quantum reflection occurs \cite{julienne1989,williams1999}. We checked that this sudden change in the interaction does not modify the quantum reflection and thus the universal collision rates defined in Refs.~\cite{julienne2011, quemener2011a}.

The full numerical formalism must take into account the coupling between rotational levels of the diatoms induced by $\hat{\mathrm{V}}_{\mathrm{dd}}$ (assuming $v_1=v_2=0$), by diagonalizing at each $R$ the Hamiltonian $\hat{\mathrm{H}} = \hat{\mathrm{H}}_{1} + \hat{\mathrm{H}}_{2} + \hat{\mathrm{V}}_{\mathrm{dd}}$, in the rotational basis $|\beta\rangle=|j_{1}m_{1}j_{2}m_{2}\rangle$, and for a given parity $p=(-1)^{j_{1}+j_{2}}$ and a given $M=m_1+m_2$. The free rotation terms $\hat{\mathrm{H}}_{i}$ of molecule $i$ only have diagonal elements equal to $B_0 j_{i}(j_{i}+1)$. The matrix elements of the dipolar Hamiltonian (\ref{eq:Vdd}) read
\begin{equation}
\langle\beta'|\hat{\mathrm{V}}_{\mathrm{dd}}|\beta\rangle = \frac{d_0^2} {R^{3}} C_{j_1010}^{j'_10} C_{j_2010}^{j'_20}
\sum_{q} A_{q}
C_{j_{1}m_{1}1q}^{j'_{1} m'_{1}}
C_{j_{2}m_{2}1-q}^{j'_{2} m'_{2}} \,,
\label{eq:Vdd-MatElem}
\end{equation}
where
\begin{equation}
A_{q} = -\frac{2}{(1+q)!(1-q)!} \sqrt{\frac{(2j_{1}+1)(2j_{2}+1)} {(2j'_{1}+1)(2j'_{2}+1)}}
\end{equation}
is a numerical factor, and $C_{a\alpha b\beta}^{c\gamma}$ are Clebsh-Gordan coefficients \cite{varshalovich1988}. Values up to $j_i=6$ for $\bar{R}>10$, $j_i=10$ for $0.25<\bar{R}<10$ and $j_i=15$ for $0.1<\bar{R}<0.25$ have been included in the calculations. In analogy with two atoms, the resulting adiabatic potential energy curves (PECs) are labeled $|M|_{g/u}^{\sigma (p)}$ where $\sigma$ is the symmetry with respect to a plane containing the intermolecular axis.

\begin{figure}
\includegraphics[width=8cm]{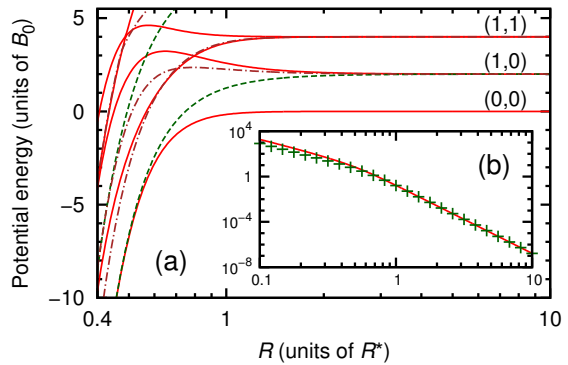}
\caption{(Color online) (a) Long-range adiabatic PECs (in scaled units) of $0_g^{+(\pm)}$ (solid lines), $0_u^{+(\pm)}$ (dashed lines), $1_u^{(\pm)}$ (dash-dotted lines) symmetries of two identical $v=0$ ground state polar diatoms. (b) the lowest $0_g^{+(+)}$ PEC in log scale (solid line: numerical; crosses: Eq.~(\ref{eq:V-2chan})).}
\label{fig:PECs}
\end{figure}

Figure \ref{fig:PECs} displays the lowest PECs calculated numerically for the $0_g^{+(\pm)}$, $0_u^{+(\pm)}$ and $1_u^{+(\pm)}$ symmetries, which will be relevant for the analysis in an electric field. Figure \ref{fig:PECs}b shows that Eq.~(\ref{eq:V-2chan}) correctly reproduces the full numerical calculations, although it underestimates the magnitude of the $R^{-3}$ interaction. For $R\gg R^*$ a one-channel description similar the one of Sec.~\ref{sec:C6} is valid. The potential energy scales as $R^{-6}$, except for the PECs connected to the channel (1,0) which scale as $R^{-3}$ due to the exchange of rotational excitation. As $R\to +\infty$ the energy of the $0_{g/u}^{+(-)}$ curves is $\bar{V}_{1,0}(R)\approx 2\pm 2/3\bar{R}^3$ (the $+$ sign is for $0_{g}^{+(-)}$), and the energy of the $1_{g/u}^{+(-)}$ curves is $\bar{V}_{1,1}(R)\approx 2\mp 1/3\bar{R}^3$. For low-$R$ values ($R\ll R^*$) the rotational levels become significantly coupled giving birth for example to potential barriers. Finally note the behavior of the lowest $0_g^{+(+)}$ and $0_u^{+(-)}$ curves which get closer and closer when $R$ decreases. This will have important consequences in the presence of an electric field.

\section{Application of an external electric field
\label{sec:Efield}}

In the T-CS the influence of an external electric field on the molecule-molecule interactions is for instance relevant when considering experimental setups using one- or two-dimensional traps for the ultracold quantum gas. The field-molecule Hamiltonian $\hat{\mathrm{W}}_i=-\hat{\vec{d}}_{i}\cdot\hat{\vec{\mathcal{E}}}$ reduces to $(-\mathcal{E}\cos\hat{\theta}_{i})$ and $(-\mathcal{E}\sin\hat{\theta}_{i} \cos\hat{\phi}_{i})$ for a parallel (along $z$, $\pi=0 \equiv \parallel$) and perpendicular (along $x$, $\pi=1 \equiv \perp$) field, respectively. Its matrix elements for molecule 1 (and vice-versa for molecule 2) coupling states with different parity $p$ and $p'$ are
\begin{eqnarray}
\langle\beta'|\hat{W}_{1}^{\pi}|\beta\rangle = - \delta_{j_{2}j'_{2}} \delta_{m_{2}m'_{2}} \sqrt{ \frac{2j_{1}+1}{2j'_{1}+1}} C_{j_{1}010}^{j'_{1}0} \Omega_{1}^{\pi}\mathcal{E}d_0
\label{eq:W-1}
\end{eqnarray}
with
\begin{equation}
\Omega_{1}^{\pi} = \frac{1}{\sqrt{2(1+\delta_{\pi0})}} \left[ C_{j_{1}m_{1}1-\pi}^{j'_{1}m'_{1}} + (-1)^{\pi}C_{j_{1}m_{1}1\pi}^{j'_{1}m'_{1}} \right].
\label{eq:Wpi-1}
\end{equation}
The dimensionless quantity $\bar{\mathcal{E}}=\mathcal{E}/\mathcal{E}^*$ holds for the electric field expressed in units of $\mathcal{E}^{*}=B_0/d_0$ (Table \ref{tab:GrandCarac}).

The electric field couples the lowest $0_{g}^{+(+)}$ state $|j_1m_1;j_2m_2\rangle=|00;00\rangle$ to symmetric superpositions of $|j_1m_1;j_2m_2\rangle$ and $|j_2m_2;j_1m_1\rangle$ states.
In the parallel case, $|00;00\rangle$ is directly coupled to the $0_{u}^{+(-)}$ states $|00;10\rangle$ and $|10;00\rangle$, while the latter are in turn directly coupled to the $0_{g}^{+(+)}$ state $|10;10\rangle$. Similarly, the perpendicular field induces a coupling between the $0_{g}^{+(+)}$  state $|00;00\rangle$ with $|00;11\rangle$, $|00;1-1\rangle$, $|11;00\rangle$, and $|1-1;00\rangle$ which combine together to form $1_u^{(-)}$ states (see Fig.~\ref{fig:PECs}a).

Therefore a perturbative calculation of the lowest PEC $\bar{V}_{00;\pi}(\bar{R})$ for $\bar{R}\gg1$ and $\bar{\mathcal{E}}\ll1$ requires:~(i) the inclusion of the $(j_1,j_2)=(0,0), (1,0), (1,1)$ channels; (ii) the third-order correction on energy, as it contains the first crossed ($\bar{R}$,$\bar{\mathcal{E}}$) contribution. In the parallel case ($\pi=0$ in Eq.~(\ref{eq:Wpi-1})), the Hamiltonian on which the perturbation theory is applied is expressed in the basis $|00;00\rangle,\,(|10;00\rangle+|00;10\rangle)/\sqrt{2},\,|10;10\rangle,(|10;00\rangle+|00;10\rangle)/\sqrt{2}$ as
\begin{equation}
\left(\begin{array}{cccc}
                           0 & -\sqrt{2/3}\bar{\mathcal{E}} &                -2/3\bar{R}^3 & -\sqrt{2}/3\bar{R}^3 \\
-\sqrt{2/3}\bar{\mathcal{E}} &             2 - 2/3\bar{R}^3 & -\sqrt{2/3}\bar{\mathcal{E}} & 0 \\
               -2/3\bar{R}^3 & -\sqrt{2/3}\bar{\mathcal{E}} &                            4 & 0 \\
        -\sqrt{2}/3\bar{R}^3 &                            0 &                            0 & 4
\end{array}\right).
\end{equation}
It gives for the lowest PEC
\begin{equation}
\bar{V}_{00;\parallel} (\bar{R},\bar{\mathcal{E}}) \approx -\frac{2\bar{\mathcal{E}}^2}{9\bar{R}^3} - \frac{1}{6\bar{R}^6}
\label{eq:V_perturb-para}
\end{equation}
where the energy of the two infinitely separated molecules has been set to 0 for each $\bar{\mathcal{E}}$.
The attractive $R^{-3}$ character of the lowest PEC, induced by the head-to-tail configuration of the two molecules, is valid only for distances such that $\bar{R}\gg \bar{\mathcal{E}}^{-2/3}$, while below this limit the vdW term is dominant.

\begin{figure}
\includegraphics[width=8cm]{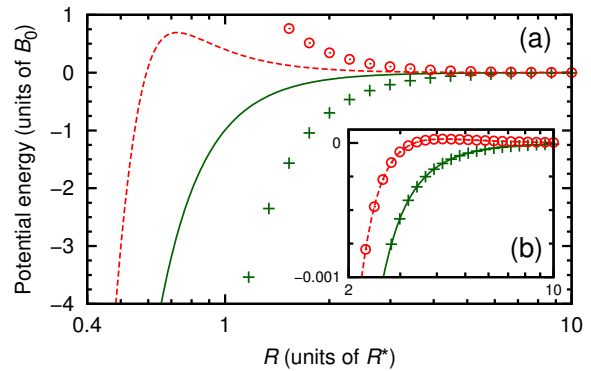}
\caption{(Color online) Long-range adiabatic PECs (in scaled units) for two identical $v=0$ ground state polar diatoms submitted to an external electric field: (a) $\mathcal{E}=5\mathcal{E}^*$; and (b) $\mathcal{E}=\mathcal{E}^{*}/5$.
The solid (dashed) lines correspond to the numerical results in a parallel (perpendicular) field. The plus signs (open circles) correspond to the analytical approximations (Eqs.~(\ref{eq:V_perturb-para}) and (\ref{eq:V_perturb-perp})) in a parallel (perpendicular) field.}
\label{fig:PECs-field}
\end{figure}

In the perpendicular case ($\pi=1$ in Eq.~(\ref{eq:Wpi-1})) a similar perturbative calculation gives for the lowest PEC
\begin{equation}
\bar{V}_{00;\bot} (\bar{R},\bar{\mathcal{E}}) \approx \frac{\bar{\mathcal{E}}^2}{9\bar{R}^3} - \frac{1}{6\bar{R}^6}.
\label{eq:V_perturb-perp}
\end{equation}
The {}``huge'' vdW term competes with a side-by-side repulsive $R^{-3}$ term, resulting in an expected potential barrier located at
\begin{equation}
\bar{R}_{b}\approx(3/\bar{\mathcal{E}}^{2})^{1/3}
\label{eq:Rbarriere}
\end{equation}
with the height
\begin{equation}
\bar{V}_{b}\approx \bar{\mathcal{E}}^{4}/54.
\label{eq:Vbarriere}
\end{equation}
In physical units the coordinates of the barrier are $R_b\approx (3B_0/\mathcal{E}^2)^{1/3}$ and $V_b\approx d_0^4\mathcal{E}^4/54B_0^3$. As $\mathcal{E}$ increases, the barrier shifts toward low $R$, but its height remains small compared to $B_0$, even for moderate fields $\bar{\mathcal{E}}\approx 1$. The position and height of the barrier, calculated for an electric field of 1kV/cm, are given in Table \ref{tab:GrandCarac} for each species. The height increases with the molecular dipole moment, so that colliding molecules like NaCs are unlikely to overcome the barrier in the ultracold regime.Figure \ref{fig:PECs-field} shows that these features are observable in the PECs obtained after the diagonalization of the molecule-molecule + molecule-field Hamiltonians (Eqs.~(\ref{eq:Vdd-MatElem}) and (\ref{eq:W-1})). For a weak field (Fig.~\ref{fig:PECs-field}b) the perturbative expressions (\ref{eq:V_perturb-para}) and (\ref{eq:V_perturb-perp}) are in very good agreement with the numerical  results. In the strong-field regime (Fig.~\ref{fig:PECs-field}a), \textit{i.e.}~outside the perturbative regime, Eq.~(\ref{eq:V_perturb-perp}) captures the essential features of the numerical results, even if the height of the potential barrier calculated numerically is smaller due to the coupling with higher levels, which are not included in the analytical estimate.

\begin{figure}
\includegraphics[width=85mm]{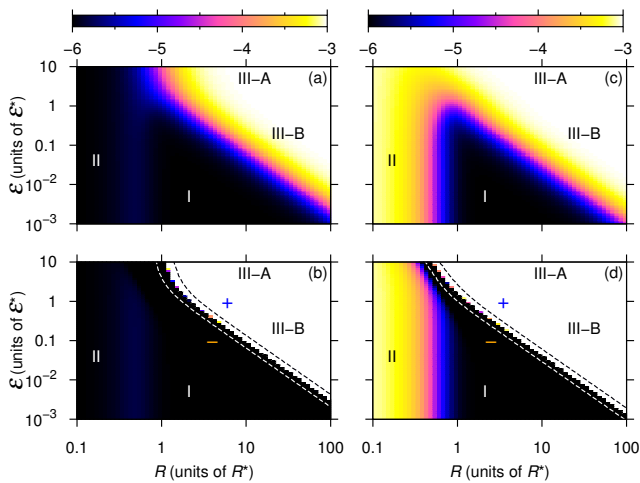}
\caption{(Color online) Leading exponent $n^*$ (Eq.~(\ref{eq:exposant})) characterizing the molecule-molecule interaction for the lowest state in parallel (upper row) and perpendicular (lower row) electric field in the T-CS as a function of ${R}$ and ${\mathcal{E}}$ in reduced units: (a)-(b) for KRb; (c)-(d) for all species but KRb and LiNa. The color scale ranges from black ($n^*=-6$) to white ($n^*=-3$). The Roman numbers correspond to regions of the ($R,\mathcal{E}$) plane characterized by different types of interactions (see text). In (b) and (d) the $+/-$ symbols refer to a repulsive/attractive interaction, while the zones between the dashed lines correspond to a change of sign of the interaction and are physically irrelevant (see text).}
\label{fig:3d-expo}
\end{figure}

The lowest PEC can be conveniently and unambiguously characterized by the leading exponent $n^{*}(\bar{R},\bar{\mathcal{E}})$ of the long-range interaction
\begin{equation}
n^{*}(\bar{R},\bar{\mathcal{E}}) = \frac{\partial\log_{10}|V_{00;\pi}(\bar{R},\bar{\mathcal{E}}))|}{\partial\log \bar{R}}\,,
\label{eq:exposant}
\end{equation}
with $n^{*}=-3$ if the interaction is of pure dipole-dipole type, and $n^{*}=-6$ if it is of pure vdW type. On Fig.~\ref{fig:3d-expo}, $n^*$ is plotted as a function of $\bar{R}$ and $\bar{\mathcal{E}}$ for parallel and perpendicular fields. Panels a and b correspond to KRb, but are also applicable to LiNa, while panels c and d correspond to the 8 other molecules. The ($\bar{R},\,\bar{\mathcal{E}}$) plane is divided in different areas labelled with Roman numbers.

In Region I the van der Waals interaction dominates the dipolar interaction even in non-vanishing fields.
Region III corresponds to the expected dipole-dipole interaction, which is attractive due to a head-to-tail approach in a parallel field, and repulsive due to a side-by-side approach in a perpendicular field. Region III spreads to the low-$R$ values with increasing fields, and its border with region I scales as $\mathcal{E}\sim R^{-3/2}$, as predicted from Eqs.~(\ref{eq:V_perturb-para}) and (\ref{eq:V_perturb-perp}). In a perpendicular field the competition between the repulsive dipole-dipole interaction and the attractive vdW interaction (represented by the {}``+'' and {}``-'' signs on Figs.~\ref{fig:3d-expo}b and \ref{fig:3d-expo}d) induces a change in the sign of the potential energy, which causes the divergence of $n^*$. On Figs.~\ref{fig:3d-expo}b and \ref{fig:3d-expo}d, the region between the dotted lines is physical irrelevance due to this divergence.

Region II of Figs.~\ref{fig:3d-expo}c and \ref{fig:3d-expo}d is characterized by a dominant dipole-dipole interaction in $R^{-3}$, as expected for the most polar molecules. Region II even exists at $\bar{\mathcal{E}}=0$ as observed on Fig.\ref{fig:PECs} and its border with Region I is field-independent. The dipole-dipole interaction strongly couples the rotational levels of the molecules, which causes their mutual orientation.
For KRb (Fig.~\ref{fig:3d-expo}(a, b)), an approximate value of $n^{*}$ is obtained by diagonalizing Eq.~(\ref{eq:Vdd-MatElem}) in physical units, and adding the diagonal contribution $-(C_6^{g-e}+C_6^e)/R^6$. It is striking to see that while Regions I and III are similar to those of the former case, the vdW interaction dominates even at short distances (Region II) even at high electric fields, due to the low value of the PEDM.

\section{Discussion: Alignment and Orientation of interacting polar molecules
\label{sec:align}}

\begin{figure}
\includegraphics[width=85mm]{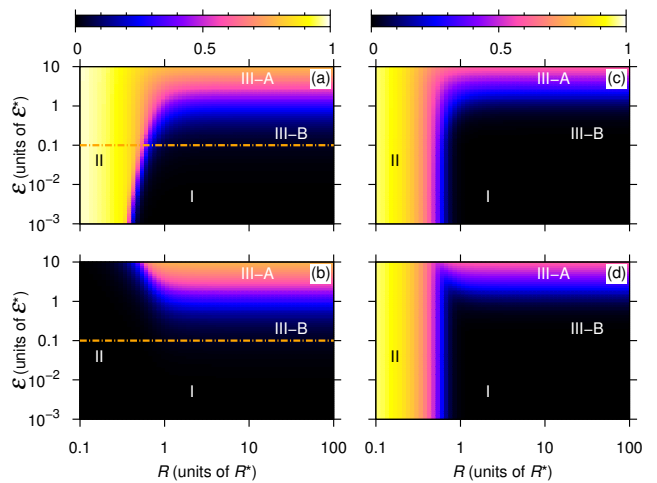}
\caption{(Color online) The molecule-molecule interaction for the lowest state in parallel (upper row) and perpendicular (lower row) electric field in the T-CS as a function of ${R}$ and ${\mathcal{E}}$ in reduced units. The color scale ranges from black (minimal values) to white (maximal values). (a)-(b) Induced dipole moment $\bar{d}_i$ of molecule $i$ ($i=1$ or 2) along the field axis  (Eq.~(\ref{eq:induc})); (c)-(d) scalar product $s$ of the two dipole moments (Eq.~(\ref{eq:align})). The Roman numbers correspond to regions of the ($R,\mathcal{E}$) plane characterized by different types of interactions (see text). The horizontal line in (a) and (b) drawn at $\mathcal{E} = \mathcal{E}^{*}/10$ locates the regime for which the variation of the induced dipole moment is discussed in the text.}
\label{fig:3d-dip}
\end{figure}

In order to understand more deeply the interaction between the eight most polar molecules, we also calculated the induced dipole moment $\bar{d}_i(\bar{R},\bar{\mathcal{E}})$ (in units of $d_0$) along the electric field axis,
\begin{equation}
\bar{d}_i(\bar{R},\bar{\mathcal{E}}) = \langle \hat{\vec{\bar{d}}}_i\cdot\hat{\vec{\bar{\mathcal{E}}}}\rangle \label{eq:induc}
\end{equation}
and the mutual alignment of the two molecules $s(\bar{R},\bar{\mathcal{E}})$, expressed as the scalar product
\begin{equation}
s(\bar{R},\bar{\mathcal{E}}) = \langle \hat{\vec{\bar{d}}}_1\cdot \hat{\vec{\bar{d}}}_2\rangle \,,
\label{eq:align}
\end{equation}
where $\langle...\rangle$ denotes the average over the eigenvector associated with the lowest PEC. The quantites $\bar{d}_i$ and $s$ are plotted on Figure \ref{fig:3d-dip} featuring the same regions as Fig.~\ref{fig:3d-expo}.

In Region I, both $\bar{d}_i$ and $s$ vanish: in a simple picture the dominant vdW interaction can be regarded as the consequence of the independent rotation of the two molecules. Although Region III displays the expected $R^{-3}$ dipolar interaction, Figure \ref{fig:3d-dip} shows that for low fields ($\bar{\mathcal{E}}<1$, Region III-B) the dipoles are not aligned along the field and average to zero, neither they are aligned against each other. The inverted situation takes place at high fields ($\bar{\mathcal{E}}>1$, Region III-A). The molecular alignment becomes significant for $\bar{\mathcal{E}}\sim 1$, \textit{i.e.}~$\mathcal{E}\sim\mathcal{E}^*$. In Tab.~\ref{tab:GrandCarac} we see that $\mathcal{E}^*$ decreases with the PEDM magnitude. It means that the molecules interacting with the strongest vdW force are actually the easiest to align along the field.

\begin{table*}
\begin{ruledtabular}
\begin{tabular}{rrrrrrrr}
           Molecule &   $C_6$ & $R_\mathrm{vdW}$ &  $R^*$ & $\mathcal{E}^*$ &  $R_b$ &              $V_b$ & $R_q$ \\
                    &  (a.u.) &           (a.u.) & (a.u.) &         (kV/cm) & (a.u.) &               (mK) & (a.u.) \\
\hline
$^{23}$Na$^{133}$Cs & 7323100 &       574 &    234 &             0.7 &    328 &               5.3  &  12 \\
$^7$Li$^{133}$Cs    & 4585400 &       497 &    178 &             2.0 &    408 &               0.3  &  10\\
$^{23}$Na$^{87}$Rb  & 1524900 &       355 &    175 &             1.3 &    294 &               0.8  &  16 \\
$^7$Li$^{87}$Rb     & 1252300 &       325 &    140 &             3.1 &    428 &               0.06 &  13 \\
$^7$Li$^{39}$K      &  570190 &       223 &    119 &             4.3 &    169 &               0.02 &  15 \\
$^{23}$Na$^{39}$K   &  561070 &       240 &    140 &             2.0 &    327 &               0.1  &  19 \\
$^{39}$K$^{133}$Cs  &  345740 &       274 &    156 &             1.0 &    226 &               0.8  &  38 \\
$^{87}$Rb$^{133}$Cs &  147260 &       236 &    148 &             0.8 &    189 &               1.0  &  60 \\
\hline
$^{39}$K$^{87}$Rb   &   15972 &       118 &     70 &             4.0 &    451 & $6{\times}10^{-4}$ & 109 \\
$^7$Li$^{23}$Na     &    3583 &        57 &     31 &            36.1 &   1130 & $6{\times}10^{-7}$ &  95 \\
\end{tabular}
\end{ruledtabular}
\caption{Various characteristic distances (see text) $R_{vdW}$, $R^*$, $R_q$, and position of the barrier ($R_b$, $V_b$) for a field $\mathcal{E}=1$~kV/cm. The $C_{6}$ coefficients (from Tab.~\ref{tab:C6}) for two identical ground-state molecules in the ($v=0, j=0$) level are recalled for the sake of clarity.}
\label{tab:GrandCarac}
\end{table*}

Figure \ref{fig:3d-dip}(c,d) shows that in Region II the molecules are aligned with respect to each other in a head-to-tail configuration ($s\approx 1$), while there is no preferential orientation ($\bar{d}_i=0$) in a perpendicular field. The large dipole-dipole interaction, compared to the Stark energy, strongly correlates the molecules to each other, whereas they are not influenced by the field. Surprisingly in a parallel field, both $s$ and $\bar{d}_i$ are close to unity, which reflects a strong molecular alignment even at small fields (see Fig.~\ref{fig:3d-dip}(a)). This comes from the degeneracy between the lowest $0_g^{+(+)}$ and $0_u^{+(-)}$ field-free PECs visible on Fig.~\ref{fig:PECs}a. A small field is sufficient to raise that degeneracy, and the lower of the resulting states is very efficiently aligned along the field (while the upper one, not shown here, is anti-aligned, \textit{i.e.} $\bar{d}_i\approx -1$).

For a moderate parallel field, say $\mathcal{E}=\mathcal{E}^*/10$ as illustrated by the dashed line on Fig.~\ref{fig:3d-dip}(a,c), the dipole moment strongly varies with $R$: it changes from $\bar{d}_i=0.18$ at $\bar{R}=0.74$ up to $\bar{d}_i=0.81$ at $\bar{R}=0.45$. For the five species (NaK, NaRb, NaCs, KCs, and RbCs) among the ten heteronuclear bialkali molecules which are found stable against ground state collisions \cite{zuchowski2010}, this feature suggests that the formation of ultracold polar tetramers could be possible by stimulated one-photon radiative association process along the lines proposed in Ref.~\cite{juarros2006}.

In contrast, such an association process cannot occur when the electric field is perpendicular to the intermolecular axis since the individual dipole moments $\bar{d}_i$ are zero. Thus we predict that this mechanism is strongly anisotropic, and that electric fields could be used to control the association of the tetramers, in a similar way than the one used to reduce the collision rate of ultracold KRb molecules in the JILA experiment.

\section{Conclusions}
\label{sec:Conclu}

In this article, we compute the long-range interactions, in free space and in an external electric field, between two identical ground-state heteronuclear bialkali molecules, for the ten species composed of Li, Na, K, Rb and Cs atoms. We find two distinct groups of molecules: on the one hand the eight ones which possess a permanent electric dipole moment larger than 1 Debye, and on the other hand, LiNa and KRb which possess a weak permanent electric dipole moment. For the eight most polar molecules, the long-range interactions are analyzed in terms of scaled intermolecular distances, electric field and energy, which are constructed from the dipole moment and rotational constant in the vibrational ground level.

The most polar molecules interact through a huge van der Waals force, up to three orders of magnitude larger than for homonuclear molecules or atoms. For distances $R$ roughly smaller than 100 a.u., this van der Waals interaction turns into a dipole-dipole interaction due to the mutual orientation of the two molecules. Although this change in behavior occurs at a distance $R^*$ of the same order of magnitude as the van der Waals length $R_\mathrm{vdW}$, we have checked that it does not alter the universal collision rates for reactive species \cite{julienne2011, quemener2011a}. In the case of LiNa and KRb, the smaller-$R$ mutual orientation does not happen because of the weak permanent electric dipole moment.

When the field is turned on, the expected dipole-dipole interaction is observed for the ten molecules, for electric field magnitudes larger than a certain threshold value. Except for LiNa this value is accessible in current ultracold experiments. In the region of mutual orientation, and for an electric field parallel to the intermolecular axis, we predict a very strong alignment of the molecules along the field axis which is due to a surprising degeneracy between two curves significantly coupled by the field. The resulting strongly $R$-varying induced dipole moment suggests the possibility of a controlled one-photon stimulated association of ultracold ground-state polar tetramers.

In Ref.~\cite{byrd2012b} the authors compute the terms of the multipolar expansion from $R^{-3}$ to $R^{-8}$, for two polar bialkali molecules with fixed distances and angles. Then those purely electronic terms are averaged on electric-field-dressed rotational levels of each molecules, for different intermolecular distances. It is shown that the quadrupole-quadrupole interaction scaling as $R^{-5}$ dominates the dipole-dipole interaction for distances lower than $R_q$ (last column of Table \ref{tab:GrandCarac}). For the six most polar molecules, $R_q$ is smaller than 20 a.u., where our model is not valid any more. For LiNa, KRb, RbCs and KCs, this interaction may play a significant role in the small-$R$ region \cite{byrd2012a, byrd2012b}; but it would require a separate study for each molecule.

Enlightening discussions with Goulven Qu\'em\'ener, Piotr {\.Z}uchowski, Jason Byrd and Robin C\^ot\'e are gratefully acknowledged. This work was supported in part by \textit{Triangle de la Physique} (contract 2008-007T-QCCM), and by the National Science Foundation (Grant No.~NSF PHY11-25915). We acknowledge the computing facility GMPCS of the LUMAT federation (FR LUMAT 2764). Laboratoire Aim\'e Cotton is a member of the \textit{Institut Francilien de Recherches sur les Atomes Froids}.

%


\end{document}